\documentclass[english, reprint,prl, twocolumn, superscriptaddress,showpacs,amsmath,amssymb]{revtex4-1}
\usepackage{lmodern}
\usepackage[T1]{fontenc}
\usepackage[latin9]{inputenc}
\usepackage[pdftex]{color}
\usepackage{amstext}
\usepackage{amssymb}
\usepackage[pdftex]{graphicx}

\makeatletter

\pdfpageheight\paperheight
\pdfpagewidth\paperwidth

\newcommand{\lyxmathsym}[1]{\ifmmode\begingroup\def\b@ld{bold}
  \text{\ifx\math@version\b@ld\bfseries\fi#1}\endgroup\else#1\fi}

\@ifundefined{textcolor}{}
{%
 \definecolor{BLACK}{gray}{0}
 \definecolor{WHITE}{gray}{1}
 \definecolor{RED}{rgb}{1,0,0}
 \definecolor{GREEN}{rgb}{0,1,0}
 \definecolor{BLUE}{rgb}{0,0,1}
 \definecolor{CYAN}{cmyk}{1,0,0,0}
 \definecolor{MAGENTA}{cmyk}{0,1,0,0}
 \definecolor{YELLOW}{cmyk}{0,0,1,0}
}

\usepackage{upgreek}
\usepackage{babel}

\newcommand{\BYOO}{Ba$_2$YOsO$_6$}
\newcommand{\BYRO}{Ba$_2$YRuO$_6$}
\newcommand{\SYRO}{Sr$_2$YRuO$_6$}
\newcommand{\SSOO}{Sr$_2$ScOsO$_6$}
\newcommand{\LNRO}{La$_2$NaRuO$_6$}

\newcommand {\TC} {$T_{\mathrm{C}}$}
\newcommand {\TN} {$T_{\mathrm{N}}$}

\newcommand{\Fmtm}{\emph{Fm}$\bar{3}$\emph{m}}

\makeatother

\begin{document}

\title{Spin-orbit coupling controlled ground state in \SSOO{}}

\author{A. E. Taylor }

\affiliation{Quantum Condensed Matter Division, Oak Ridge National Laboratory,
Oak Ridge, Tennessee 37831, USA}

\author{R. Morrow}

\affiliation{Department of Chemistry, The Ohio State University, Columbus, Ohio
43210-1185, USA}

\author{R. S. Fishman }

\affiliation{Materials Science and Technology Division, Oak Ridge National Laboratory,
Oak Ridge, Tennessee 37831, USA}

\author{S. Calder}

\affiliation{Quantum Condensed Matter Division, Oak Ridge National Laboratory,
Oak Ridge, Tennessee 37831, USA}

\author{A.I. Kolesnikov}

\affiliation{Chemical and Engineering Materials Division, Oak Ridge National Laboratory,
Oak Ridge, Tennessee 37831, USA}

\author{M. D. Lumsden}

\affiliation{Quantum Condensed Matter Division, Oak Ridge National Laboratory,
Oak Ridge, Tennessee 37831, USA}

\author{P. M. Woodward}

\affiliation{Department of Chemistry, The Ohio State University, Columbus, Ohio
43210-1185, USA}

\author{A. D. Christianson}

\affiliation{Quantum Condensed Matter Division, Oak Ridge National Laboratory, Oak Ridge, Tennessee 37831, USA}
\affiliation{Department of Physics and Astronomy, The University of Tennessee, Knoxville, TN 37996, USA}


\begin{abstract}

We report neutron scattering experiments which reveal a large spin gap in the magnetic excitation spectrum of weakly-monoclinic double perovskite \SSOO{}. The spin gap is demonstrative of appreciable spin-orbit-induced anisotropy, despite nominally orbitally-quenched 5\emph{d}$^{3}$ Os$^{5+}$ ions. The system is successfully modeled including nearest neighbor interactions in a Heisenberg Hamiltonian with exchange anisotropy. We find that the presence of the spin-orbit-induced anisotropy is essential for the realization of the type I antiferromagnetic ground state. This demonstrates that physics beyond the LS or JJ coupling limits plays an active role in determining the collective properties of $4d^{3}$ and $5d^{3}$ systems, and that theoretical treatments must include spin-orbit coupling.

\end{abstract}
\maketitle

The role of spin-orbit coupling (SOC) in 4$d$ and 5$d$ transition metal oxides is relatively poorly understood outside of the $LS$ and $JJ$ coupling limits. 
The need to understand the intermediate regime is typified by the diverse range of properties found in double perovskites (DPs) containing $4d$ and $5d$ ions, including high-temperature half-metallic ferrimagnetism~\cite{kobayashi_room-temperature_1998,kobayashi_intergrain_1999}, structurally selective magnetic states~\cite{feng_high-temperature_2014,morrow_probing_2014,morrow_effects_2015}, complex geometric frustration~\cite{aczel_frustration_2013,aharen_magnetic_2009,bernardo_magnetic_2015,carlo_spin_2013,e._v._kuzmin_effect_2003,kermarrec_frustrated_2015},  and Mott insulating states~\cite{erickson_ferromagnetism_2007,gangopadhyay_spin-orbit_2015, meetei_theory_2013}. Whilst the complex array of ground states has generated a great deal of interest, the interaction mechanisms controlling them remain undetermined. 

\begin{figure}[t]
\includegraphics[width=0.98\columnwidth]{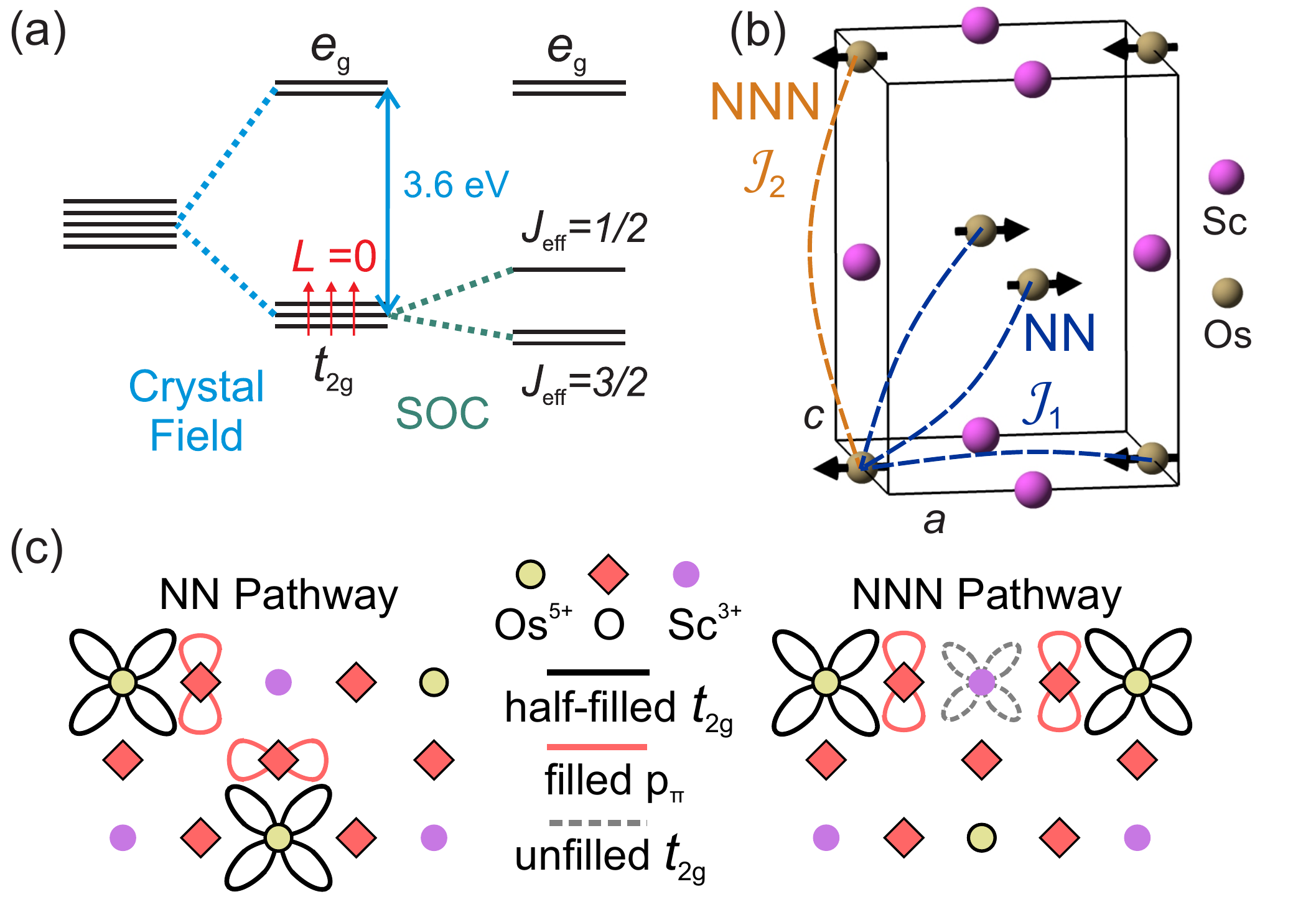}
\protect\caption{\label{levels_struct} (a) Schematic of the energy
levels for Os$^{5+}$ in an octahedral environment. 
$t_{2\mathrm{g}}$--$e_{\mathrm{g}}$ spliting of 3.6$\,$eV determined by
x-ray absorption spectroscopy~\cite{choy_study_1998}. In the strong
SOC limit the $t_{2\mathrm{g}}$ level can be further split into $J_{\mathrm{eff}}=\frac{1}{2}$
and $\frac{3}{2}$ levels. Nominally the Os$^{5+}$ ion is in the
$LS$ coupling limit and an $L=0$ state results. (b) \SSOO{} magnetic
structure, with moments depicted along $a$.
One $P2_{1}/n$ unit cell is shown, with O and Sr ions omitted for
clarity. Dashed lines show examples of the NN ($\times$12)
$\mathcal{J}_{1}$ and NNN ($\times$6) $\mathcal{J}_{2}$
exchanges. (c)~Schematic of the relevant orbitals for NN and NNN exchange pathways, assuming formal valence states.}
\end{figure}

For DPs hosting 4\emph{$d^{3}$} or 5\emph{$d^{3}$} ions, the role of SOC is particularly unclear. 
There exists dispute between different theories describing SOC and its influence on the interactions~\cite{middey_route_2012,matsuura_effect_2013,e._v._kuzmin_effect_2003,chen_spin-orbit_2011,meetei_theory_2013, das_origin_2011,sanyal_ferromagnetism_2014,samanta_half-metallic_2015}. To first order, $d{}^{3}$ ions in an octahedral environment are expected to be orbitally quenched, Fig.~\ref{levels_struct}(a)~\cite{chen_spin-orbit_2011,carlo_spin_2013}, yet there is mounting evidence that SOC has considerable influence~\cite{aczel_exotic_2014,aczel_frustration_2013,kermarrec_frustrated_2015,nilsen_diffuse_2015,taylor_magnetic_2015}. 
This has been demonstrated by the presence of $\sim$2--18\,meV gaps in the magnetic excitation spectra of
\BYRO{}, \LNRO{} and \BYOO{}~\cite{carlo_spin_2013,kermarrec_frustrated_2015,aczel_exotic_2014}. Such large gaps, on the same energy scale as the \TN{}s, implies a departure from an orbital singlet, and raises the question of how SOC manifests in the collective properties. 


Beyond a fundamental interest in the influence of SOC, it is vital to determine the sign and strength of exchange interactions between 5$d$ ions in order to understand the magnetism of many DPs, including the exceptionally high $T_\mathrm{C} = 725\,$K seen in Sr$_2$CrOsO$_6$~\cite{krockenberger_neutron_2007,krockenberger_sr2croso6:_2007}.  Investigations of Sr$_2$CrOsO$_6$ and related materials show that exchange interactions between Os$^{5+}$ ions cannot be neglected~\cite{meetei_theory_2013, das_origin_2011, sanyal_ferromagnetism_2014, feng_high-temperature_2014, paul_lattice_2013, taylor_magnetic_2015,samanta_half-metallic_2015}.  However, the strong coupling between Cr$^{3+}$ and Os$^{5+}$ ions makes it difficult to measure the strength of the Os-Os coupling. Additionally, there is a lack of agreement regarding the mechanism that stabilizes type I antiferromagnetic (AFM) order on the face-centered-cubic (FCC) lattice of $B^\prime$ ions in $A_{2}BB^\prime$O$_6$ DPs, where $B$ is diamagnetic, and $B^\prime$ is either Ru$^{5+}$ (4$d^3$) or Os$^{5+}$ (5$d^3$)~\cite{e._v._kuzmin_effect_2003,kermarrec_frustrated_2015}. Most attempts to determine the exchange interactions in these systems have been limited to theoretical models not directly related to measurements, with conflicting results~\cite{meetei_theory_2013, kanungo_textitab_2014,hou_lattice-distortion_2015,wang_first-principle_2014,e._v._kuzmin_effect_2003}. Therefore, to understand the underlying behavior, it is desirable to obtain the interactions experimentally.
%


To access Os$^{5+}$ ion interactions experimentally, we investigate \SSOO{}. It is the single-magnetic-ion analogue of Sr$_2$CrOsO$_6$, therefore all magnetic interactions result solely from the frustrated quasi-FCC Os$^{5+}$ lattice. Despite this, \SSOO{} hosts a remarkably high \TN{} (92 K) for a single-magnetic-ion DP~\cite{paul_magnetically_2015,taylor_magnetic_2015,yuan_high-pressure_2015}. It is therefore a model system for investigating the role of the Os$^{5+}$ $5d^3$ magnetic interactions in a high transition temperature material. 


We present the inelastic neutron scattering (INS) spectrum of \SSOO{}, and find a large spin gap below \TN{}. 
A Heisenberg Hamiltonian with anisotropic exchange terms
is considered. We find that over a large parameter space, the solution which best describes the data is one with the isotropic nearest-neighbor (NN) term $J_{1}=-4.4$\,meV, and negligible next-nearest-neighbor (NNN) interactions. The success
of the model reveals that anisotropy is essential to selection of the type I AFM ground state. This suggests that SOC  within the 5$d^{3}$ manifold, along with strong Os-O hybridization, promotes a high \TN{} in this otherwise frustrated material. Therefore, it is NN interactions combined with SOC-induced anisotropy that are key to the collective behavior realized in \SSOO{}, and related $4d^{3}$ and $5d^{3}$ systems. This demonstrates that SOC must be included in theoretical treatments of these materials.

A 16.5\,g polycrystalline sample of \SSOO{} was used for INS experiments
on SEQUOIA at the Spallation Neutron Source
at Oak Ridge National Laboratory~\cite{granroth_sequoia:_2010}, see Supplemental Material (SM)~\cite{supplementary} for full details.
The structural and magnetic properties of the same sample were reported in Ref.~\cite{taylor_magnetic_2015}, finding space group $P2_1/n$ with $a = 5.6398(2)\,\mathrm{\AA}$, $b= 5.6373(2)\,\mathrm{\AA}$, 
$c = 7.9884(3)\,\mathrm{\AA}$ and $\beta = 90.219(2)^{\circ}$ at 5\,K, and $T_\mathrm{N}=92\,$K.

\begin{figure}
\includegraphics[width=1\columnwidth]{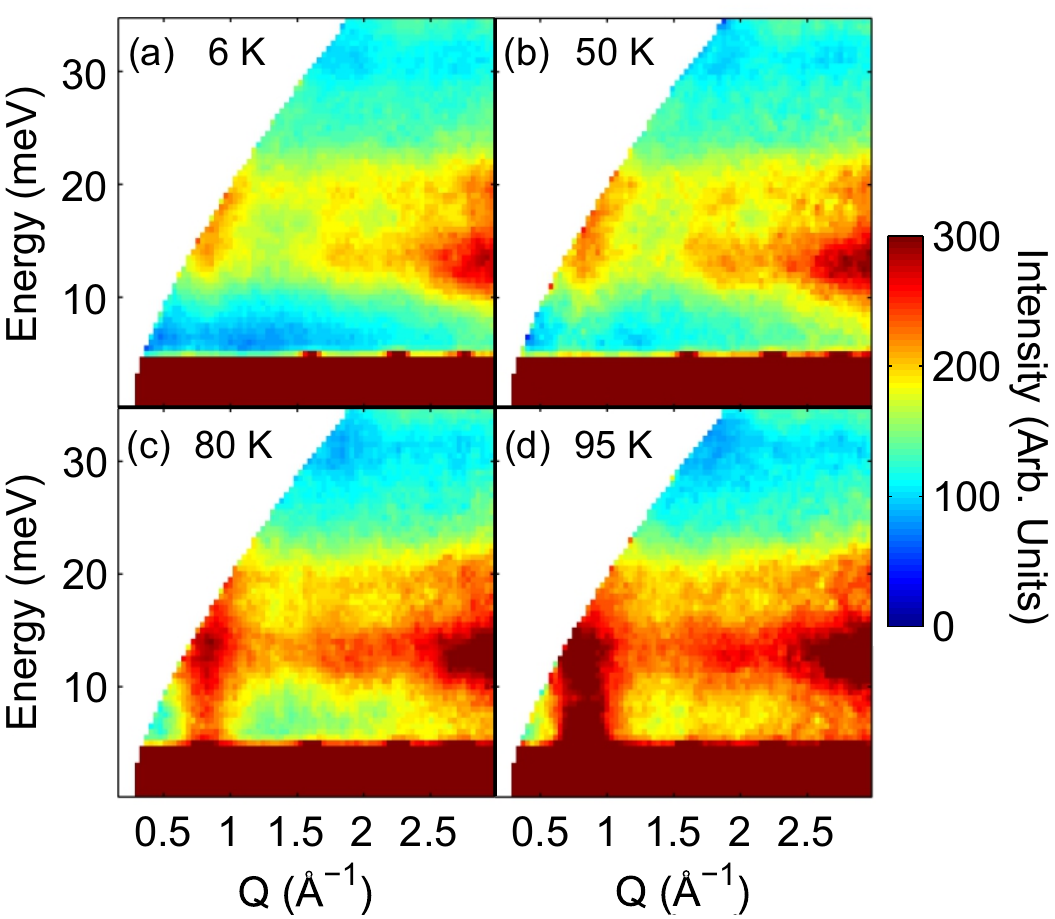}
\protect\caption{\label{Slices} $E_{\mathrm{i}}=60\,$meV  neutron scattering intensity maps for $95\,\mathrm{K}\gtrapprox T_\mathrm{N}$, and $T<$\TN{} of 80, 50 and 6\,K. }
\end{figure}

Measured INS spectra are shown in Fig.~\ref{Slices}. There is a pronounced
change in the spectrum at low neutron momentum transfer ($Q$) upon
crossing \TN{}. This behavior is reminiscent of
the observed gap development below \TN{} in other
single magnetic ion 4\emph{d}$^{3}$ and 5\emph{d}$^{3}$ DPs~\cite{aczel_exotic_2014,carlo_spin_2013,kermarrec_frustrated_2015}.
The higher $Q$ scattering, which changes only in intensity with temperature, is identified as phonon scattering. 

\begin{figure}
\includegraphics[width=1\columnwidth]{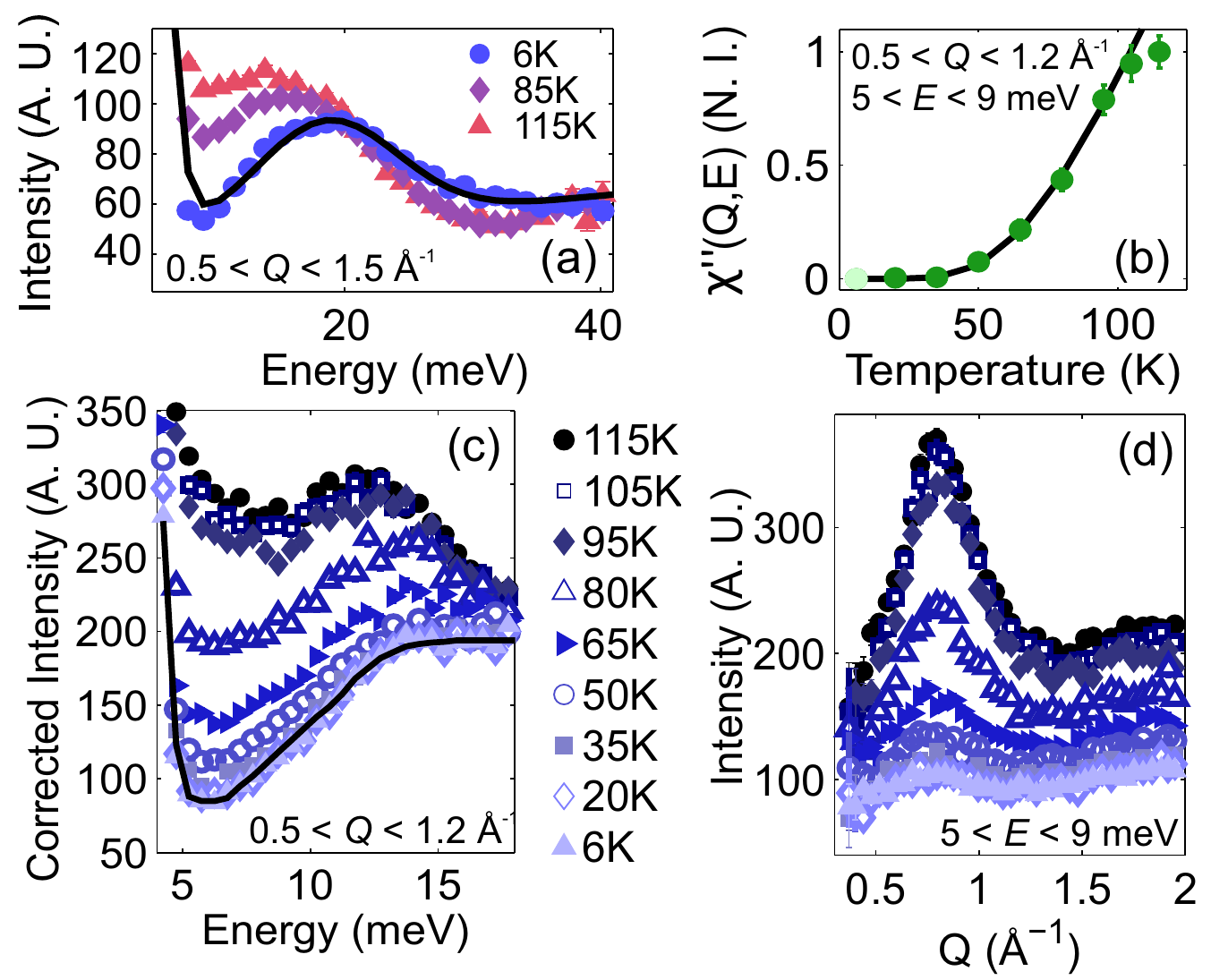}
\protect\caption{\label{INS_cuts} (a) Constant-$Q$ cuts from $E_{\mathrm{i}}=120\,$meV
data. Solid line is the result of fitting Gaussians to the elastic
line and to the inelastic magnetic signal at 6\,K. A. U. stands for
arbitrary units. (b) $\chi^{\prime\prime}(T)$ at fixed $Q$ and $E$, with an exponential, $\chi^{\prime\prime}(T)\propto\exp(-\Delta/k\mathrm{_{B}T})$,
fit to the $T<T_{\mathrm{N}}$ data. N.I. stands for normalized intensity. (c) Constant-$Q$ cuts from
$E_{\mathrm{i}}=60\,$meV data, which have been corrected for the
Bose factor. Solid line is a guide to the eye. (d) Constant-$E$ cuts
from $E_{\mathrm{i}}=60\,$meV data. In all panels, errorbars are sometimes smaller than the symbols. }
\end{figure}

The detailed $(Q,E)$-space structure and temperature dependence of
the scattering is presented in Fig.~\ref{INS_cuts}. Figure~\ref{INS_cuts}(a)
demonstrates that intensity is distributed to higher energies
at low temperatures, as expected from a gap opening. The
peak of the scattering intensity at 6\,K is at $\eta=19(2)$\,meV. This
compares to previous observations,  which have been used as a magnitude estimate for the gap, of
$\eta=18(2)\,$meV in \BYOO{} ($T_\mathrm{N}$=69\,K), $\eta\approx5\,$meV in \BYRO{}  ($T_\mathrm{N}$=36\,K) and
$\eta\approx2.75$\,meV in \LNRO{} ( $T_\mathrm{N}$=15\,K)~\cite{kermarrec_frustrated_2015,carlo_spin_2013,aczel_exotic_2014}.
This generally supports a picture of gap energy scale varying with
\TN{}. Figure~\ref{INS_cuts}(c) presents data that has been corrected for the Bose thermal population factor,
$[1-\exp(-E/k_{\mathrm{B}}T)]^{-1}$. The sharp drop in intensity at low $E$ below \TN{} demonstrates the opening of the
gap. 

Constant-$E$ cuts averaged from 5 to 9\,meV show scattering centered around AFM ordering wavevector
$|\mathbf{Q}_{(001)}|\approx0.8\,\mathrm{\AA^{-1}}$, Fig.~\ref{INS_cuts}(d), with some asymmetry
in the lineshape resulting from $|\mathbf{Q}_{(100)/(010)}|\approx1.1\,\mathrm{\AA^{-1}}$  fluctuations.
To track the relative strength of the fluctuations we
extract the dynamic susceptibility, $\chi^{\prime\prime}(T)$, for
fixed range $5<E<9\,$meV and $0.5<Q<1.2\,\mathrm{\AA}^{-1}$ via the same
method as Ref.~\cite{kermarrec_frustrated_2015} (see also SM~\cite{supplementary}).
The opening of a gap below \TN{} is again indicated, Fig.~\ref{INS_cuts}(b), by the reduction in $\chi^{\prime\prime}(T)$ evaluated at low-energy.

We investigate a model Heisenberg Hamiltonian with anisotropic exchange terms. The results we present here include only NN terms, $\mathcal{J}_1$, (see Fig.~\ref{levels_struct}(b)) because the NNN terms, $\mathcal{J}_2$, are dramatically suppressed (estimated as $J_2\leq 0.01 J_1$ in Ref.~\cite{e._v._kuzmin_effect_2003}), as discussed below. We tested this assumption by seeking solutions over a wide range of parameter space with $\mathcal{J}_2\neq 0$, see SM~\cite{supplementary}, but found that  $\mathcal{J}_2=0$ provided the best description of the experimental INS data. 

The model is parametrized with an isotropic term, $J_1$, which is decoupled from the physical origin of the spin gap, and an
exchange anisotropy term, $K_1$, to account for the gap. Unlike isotropic exchange terms, anisotropic exchange terms only couple to a particular component of spin, e.g. $S_{x}$. $x$ represents the direction of spin alignment. We assume that the exchange interactions are unaffected by the weak monoclinic distortion, justified by two considerations:  first, the distortion is much smaller than found in $d^3$ systems in which the distortion is reported to affect the physical properties~\cite{aczel_frustration_2013, retuerto_switching_2007, supplementary}. Secondly, the properties of the closely related cubic compound \BYOO{} are remarkably similar to \SSOO{}~\cite{kermarrec_frustrated_2015}. 
The Hamiltonian is therefore
$$ \mathcal{H} = -\sum_{\mathrm{NN}} \mathcal{J}_{1}^{\alpha\beta} S_{i\alpha} S_{j\beta} = -\sum_{\mathrm{NN}}\left(J{_1}\mathbf{S}_i \cdot \mathbf{S}_j  + K{_1}S_{ix}S_{jx}\right).$$
$J_{1}$ and $K_{1}$ are defined such that positive values are ferromagnetic (FM) and negative values are
AFM. The exchange parameters scale inversely with spin, with $s=0.8$ determined from neutron diffraction~\cite{taylor_magnetic_2015}~\footnote{From $m = 1.6(1)\,\mu_\mathrm{B}$ assuming a g-factor of 2, deemed reasonable since density functional theory finds the ratio of spin to orbital moments is $\sim$15~\cite{taylor_magnetic_2015}.}.

To accurately reproduce the INS data from \SSOO{}, we use the bottom and top of the spin wave band, $\Delta=12\,$meV  and $\Gamma=40\,$meV, respectively, as conditions to determine the parameters $J_1$ and $K_1$. $\Delta$ was determined by inspection of the 6$\,$K data in Fig.~\ref{INS_cuts}(c), in which the increasing intensity begins to saturate at $E\approx12\,$meV. $\Gamma$ was determined by inspection of broad constant-$Q$ cuts from the $E_{\mathrm{i}}=120\,$meV data (see SM Fig.~S2~\cite{supplementary}), designed to capture all magnetic scattering up to high energies, in which 6$\,$K and 115$\,$K cuts converge at 40\,meV. An additional constraint for the local stability of the ground state, depicted in Fig.~\ref{levels_struct}(b), is that the spin-wave frequencies are real throughout the magnetic Brillouin zone.  Utilizing this model, we find the solution $J_1=-4.4\,$meV and $K_1= - 3.8\,$meV. This gives a mean-field transition temperature of 181\,K, two times greater than the measured \TN{}. This is reasonable, as calculated mean-field temperatures are generally expected to exceed measured values~\cite{mattis_theory_1965}, and the Curie-Weiss constant for this compound, $\Theta = -677$\,K~\cite{taylor_magnetic_2015}, is also far greater than $T_\mathrm{N}=92$\,K.  

\begin{figure}
\includegraphics[width=1\columnwidth]{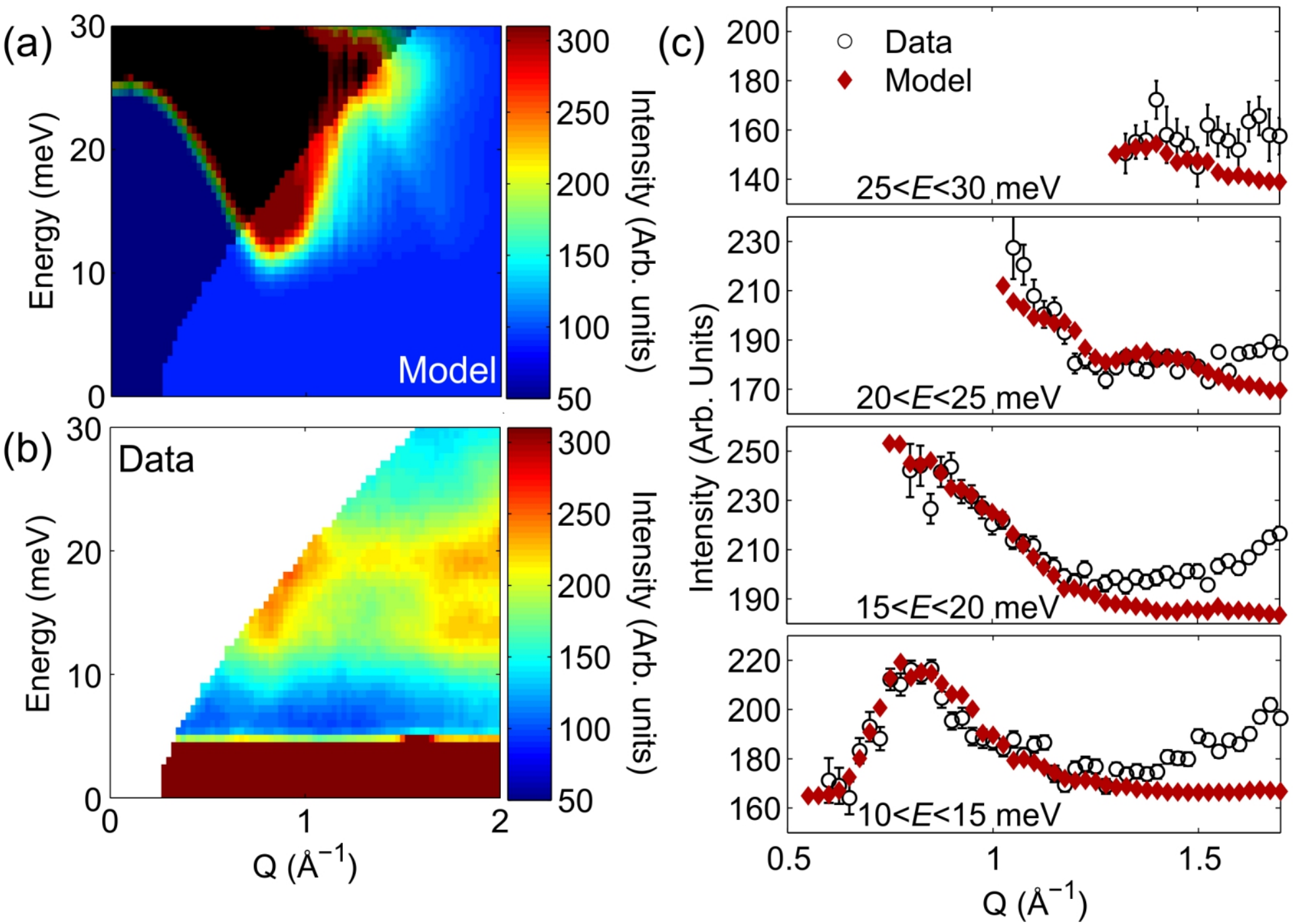}
\protect\caption{\label{Calc} (a) Simulated spin-wave spectra. Modeled using linear spin-wave theory~\cite{toth_linear_2015}, with powder averaging performed by sampling $10{}^{4}$ random points in reciprocal space. Gaussian energy broadening of  4$\,$meV is applied as an approximation to instrument resolution at $E_i=60\,$meV, estimated from the full width at half maximum of the incoherent part of the elastic line in the data. (b) The equivalent data collected at $T=6\,$K.
The intensity at high $Q$ in the data is due to phonon scattering, which is not included in the model. The shaded region in the calculations indicates the region of $(Q,E)$ space which is inaccessible in the experiment. (c) Constant-energy cuts through the calculation and data. A global scale factor has been used for the calculation, and a flat background applied for each cut.}
\end{figure}

The simulated powder-averaged INS cross section $S(Q,E)$ for  $J_1=-4.4\,$meV and $K_1= - 3.8\,$meV is compared  to the low-temperature data in Fig.~\ref{Calc}, and we find good agreement. An overview is provided by colormaps in Fig.~\ref{Calc}(a) and (b), and a more detailed comparison is given by constant-energy cuts in Fig.~\ref{Calc}(c). Note that this solution is equivalent to a single-ion anisotropy model with $J_1=-4.4\,$meV and $D=7.5$\,meV. 


Although SOC has been noted as the origin of the spin gap in $5d$ 
DPs~\cite{aczel_exotic_2014,kermarrec_frustrated_2015}, the underlying mechanism by which it acts to produce the gap remains an open question. 
In general, the possible mechanisms in a three dimensional system are Dzyaloshinsky-Moriya (DM)
interactions, single-ion anisotropy, and exchange anisotropy, all
of which are induced by SOC.
There are two observations which favor dismissal of the DM interaction
as the origin of the gap: (i) the highly symmetric cubic or close-to-cubic
crystal structures in which the gap has been observed (space group \Fmtm{} has inversion symmetry at the Os site, \emph{P}2$_1$/\emph{n} does not)
and (ii) the type I collinear AFM structure common to several DPs
including \SSOO{} -- two perpendicular DM interactions would be
required to produce a gap, but would favor a non-collinear spin
state. 

We also expect that single-ion anisotropy is negligible,
because it is dramatically suppressed for the orbitally suppressed
$d^{3}$ configuration, and the 3.6\,eV $t_{2\mathrm{g}}$ to $e_{\mathrm{g}}$
splitting in \SSOO{}~\cite{choy_study_1998} means that the excited
state perturbations are minimal~\cite{khomskii_transition_2014}.
This is supported by the experimental observation that no gap emerges
in La$_{2}$NaOsO$_{6}$ which only displays short-range order, whereas
a gap is observed in long-ranged-ordered sister-compound La$_{2}$NaRuO$_{6}$\textcolor{black}{~\cite{aczel_exotic_2014}}. A single-ion term, being a local effect, would not be sensitive to short- versus long-range order, and would emerge in the short-range
ordered state. Therefore, exchange anisotropy is the most-likely explanation for the gap in
$4d^{3}$ and $5d^{3}$ DPs. 
Independent of the gap's origin, the determination that $J_{1}\approx-4.4\,$meV and $J_{2}$ is negligibly small, has significant consequences. 


There is dispute in the literature over the strength of long-range interactions in $d^3$ DPs, and the origin of type I AFM order in $4d$ and $5d$ single-magnetic-ion DPs. Competition between type I and type III order results in frustration on the (quasi-)FCC lattice of Os/Ru ions. Theoretical studies found that type I order can be stabilized either by a FM $J_{2}$ in an isotropic (i.e. $K_1=0$) Heisenberg Hamiltonian, or by some form of anisotropy~\cite{e._v._kuzmin_effect_2003}. Nilsen \emph{et al.}~\cite{nilsen_diffuse_2015} attempted to extract the interactions in \BYRO{} via Reverse Monte Carlo (RMC) analysis of diffuse neutron scattering, and found large interactions beyond NN, with $|J_2| \approx \frac{1}{2} |J_1|$. However, by use of an isotropic Heisenberg Hamiltonian, their analysis implicitly assumed significant long-range interactions to stabilize the correct ground state, and, as they point out, could not distinguish from an anisotropy-based model. We have found that, in-fact, an NN-only exchange model with significant SOC-induced anisotropy provides the best description of the INS spectrum for \SSOO{}. 

Our result can be rationalized based on the superexchange pathways present, illustrated in Fig.~\ref{levels_struct}(c). The NN Os-O-O-Os superexchange pathway is anticipated to be strongly AFM due to the half-filled Os$^{5+}$ $t_{2g}$ levels~\cite{goodenough_theory_1955,kanamori_superexchange_1959}. Direct $t_{2g}$-$t_{2g}$ overlap is also an AFM NN contribution. The NNN pathway, however, relies on overlap with empty Sc$^{3+}$ $t_{2g}$ orbitals, and was estimated as $J_2\leq 0.01 J_1$ in Ref.~\cite{e._v._kuzmin_effect_2003}, consistent with our result. 

This analysis is, however, at odds with attempts to model the exchange interactions in 3$d^x$-5$d^3$ DPs, including Sr$_2$CrOsO$_6$, using density functional theory~\cite{das_origin_2011,kanungo_textitab_2014,hou_lattice-distortion_2015,wang_first-principle_2014}. Studies estimated $|J_2|$ in the range 1.9--24\,meV (for $s=0.8\,$meV), but did not consider the anisotropy terms (single-ion or exchange anisotropy) reported here, despite mentioning the likely frustration of Os$^{5+}$ ions. Therefore, much like the modeling of Ba$_2$YRuO$_6$ via RMC, the longer-range interactions may have been implicitly forced to have large values. This is particularly relevant in Sr$_2$CrOsO$_6$, in which both magnetic ions have $d^3$ configuration, therefore unlike (Ca,Sr)$_2$FeOsO$_6$ no occupied $e_g$ orbital pathways contribute to longer-range interactions~\cite{morrow_probing_2014,veiga_fragility_2015}. Anisotropy could therefore have a major influence in Sr$_2$CrOsO$_6$, and further calculations including anisotropy terms would be illuminating. Similar calculations for Sr$_2$ScOsO$_6$ will be directly constrained by the size of the observed gap and by $J_1\approx-4.4\,$meV, independent of the gap's origin.

As anisotropy is essential in stabilizing the AFM order in \SSOO{}, it should also be relevant in type I \BYOO{}, \BYRO{} and \SYRO{}~\cite{battle_crystal_1989,aharen_magnetic_2009,kermarrec_frustrated_2015,battle_crystal_1984}. Diffraction experiments found no structural distortion (\BYOO{} and \BYRO{}), or a small monoclinic distortion (\SYRO{}), therefore the same interaction pathways as for \SSOO{} are applicable. Although exchange/single-ion anisotropies are formally absent (to 2nd order) in a cubic system~\cite{khomskii_transition_2014}, the type I order should coincide with a distortion via magneto-elastic coupling in \BYOO{} and \BYRO{}. Although this structural distortion, if present, is outside the range of detection of present diffraction experiments, it would allow anisotropy to enter the Hamiltonian. Anisotropy has been directly observed via spin-gaps in both these materials~\cite{carlo_spin_2013,kermarrec_frustrated_2015}. We therefore propose that in all these systems, SOC is essential in determining the magnetic ground state.

%
%
%


Amongst these materials, \SSOO{} boasts the highest \TN{}. As has previously been noted, large Os-O hybridization is an important factor in heightened \TN{}s~\cite{das_origin_2011, taylor_magnetic_2015}. Our results suggest that, by promoting selection of a particular ground state and relieving frustration, Os$^{5+}$ SOC also acts to enhance \TN{} in \SSOO{}. This notion is supported by the trend in gap size with \TN{} across the measured compounds, and by the observation that $3d$ transition metal DPs have lower \TN{}s and usually favor a different, Type II, ground state~\cite{vasala_a2bbo6_2015}.


It is also informative to compare Sr$_2$ScOsO$_6$ to the equivalent 5$d^2$ systems Sr$_2$MgOsO$_6$~\cite{yuan_high-pressure_2015} and Sr$_2$ScReO$_6$~\cite{kato_structural_2004, winkler_magnetism_2009}. We expect 5$d^2$ ions to have a smaller magnetic moment~\cite{thompson_long-range_2014}, and reduced Os-O-O-Os AFM superexchange as the $t_{2g}$ levels are not half-filled. This results in a lower AFM energy scale, but unquenched SOC, which will promote a high \TN{} compared to that AFM energy scale if the SOC promotion of \TN{} is correct. Both these expectations are met - compared to \SSOO{} these compounds have lower inherent energy scales as indicated by their Curie Weiss constants, but have \TN{}s of 105\,K and 75\,K, comparable to that of \SSOO{}. Therefore SOC has an important role in high-\TN{} DPs beyond the 5$d^3$ case.

In conclusion, by modeling the magnetic excitation spectrum of archetypal system \SSOO{}, we have extracted the exchange parameters resulting from Os$^{5+}$ ion interactions. The presence of a large spin gap demonstrates that SOC is significant, i.e. the 5$d^3$ ions deviate from the nominal orbital singlet expected from LS coupling. We find that only NN interactions are significant, and as a consequence, SOC-induced anisotropy governs the magnetic state in this otherwise frustrated system, and assists in promoting a high \TN{}. This demonstrates that the interplay of NN interactions with anisotropy should be considered for the collective properties of high-\TC{} $5d^3$ systems, particularly Sr$_2$CrOsO$_6$.

\begin{acknowledgments}
The authors gratefully acknowledge M.~B.~Stone, S.~E.~Nagler and B.~D.~Gaulin for useful discussions. The research at Oak Ridge National Laboratory's Spallation Neutron Source was supported by the Scientific User Facilities
Division, Office of Basic Energy Sciences, U.S. Department of Energy (DOE).  Support for a portion of this research was provided by the Center for Emergent Materials an NSF Materials Research Science and Engineering Center (DMR-1420451). Research by RF sponsored by the DOE, Office of Science, Basic Energy Sciences, Materials Sciences and Engineering Division. 

This manuscript has been authored by UT-Battelle, LLC under Contract No. DE-AC05-00OR22725 with the U.S. Department of Energy.  The United States Government retains and the publisher, by accepting the article for publication, acknowledges that the United States Government retains a non-exclusive, paid-up, irrevocable, world-wide license to publish or reproduce the published form of this manuscript, or allow others to do so, for United States Government purposes.  The Department of Energy will provide public access to these results of federally sponsored research in accordance with the DOE Public Access Plan (http://energy.gov/downloads/doe-public-access-plan).
\end{acknowledgments}

\bibliographystyle{apsrev4-1}
\bibliography{Sr2ScOsO6_INS_bib}

 \clearpage

 \addtolength{\oddsidemargin}{-0.75in}
 \addtolength{\evensidemargin}{-0.75in}
 \addtolength{\topmargin}{-0.725in}

 \newcommand{\addpage}[1] {
     \begin{figure*}
       \includegraphics[width=8.5in,page=#1]{supplementary.pdf}
     \end{figure*}
 }

 \addpage{1}
 \addpage{2}
 \addpage{3}
 \addpage{4}

\end{document}